\documentstyle[mprocl]{article}

\def\be{\begin{equation}}
\def\ee{\end{equation}}
\def\bea{\begin{eqnarray}}
\def\eea{\end{eqnarray}}

\def\lsim{\mathrel{\rlap {\raise.5ex\hbox{$ < $}}
{\lower.5ex\hbox{$\sim$}}}}
\def\gsim{\mathrel{\rlap {\raise.5ex\hbox{$ > $}}
{\lower.5ex\hbox{$\sim$}}}}

\newcommand{\pr}{\paragraph{}}
\newcommand{\nn}{\nonumber}

\def\gappeq{\mathrel{\rlap {\raise.5ex\hbox{$>$}}
{\lower.5ex\hbox{$\sim$}}}}

\def\lappeq{\mathrel{\rlap{\raise.5ex\hbox{$<$}}
{\lower.5ex\hbox{$\sim$}}}}

\begin{document}

\title{EVIDENCE FOR
DEVIATIONS FROM FERMI-LIQUID BEHAVIOUR IN
(2+1)-DIMENSIONAL QUANTUM ELECTRODYNAMICS AND THE
NORMAL PHASE OF
HIGH-$T_c$ SUPERCONDUCTORS }

\author{ I.J.R. AITCHISON, N.E. MAVROMATOS~\footnote{Invited speaker
at the 4-th Chia meeting on `common trends in particle
and condensed matter physics', Chia-Laguna, Sardegna, Italy,
September 1995 }}

\address{University of Oxford,
Department of Physics, Theoretical Physics, \\
1 Keble Road, Oxford OX1 3NP, United Kingdom}


\maketitle\abstracts{
We provide evidence
that the gauge-fermion interaction
in multiflavour quantum electrodynamics in $(2 + 1)$-dimensions
is responsible for non-fermi liquid behaviour in the infrared,
in the sense of leading to the existence of a non-trivial
(quasi) fixed point (cross-over) that lies
between the trivial fixed point (at infinite momenta)
and the region where dynamical symmetry breaking
and mass generation occurs. This quasi-fixed point
structure implies slowly varying, rather than fixed, couplings
in the intermediate regime of momenta, a situation which
resembles that of (four-dimensional) `walking technicolour' models of
particle physics.
The
inclusion of wave-function renormalization
yields marginal $O(1/N)$-corrections
to the `bulk' non-fermi liquid behaviour
caused by the gauge interaction in the limit of
infinite flavour number.
At low temperatures there appear to be
logarithmic scaling violations of the linear resistivity
of the system of order $O(1/N)$.
Connection with
the anomalous normal-state properties of certain
condensed matter systems relevant for
high-temperature
superconductivity is briefly discussed. The
relevance of the large (flavour) $N$ expansion
to the fermi-liquid problem is emphasized.}

\section{Introduction}
\pr
One of the most striking phenomena associated
with the novel high-temperature superconductors
is their {\it abnormal} normal-state properties.
In particular, these substances are known to exhibit
deviations from the known Fermi-liquid behaviour,
which are remarkably stable with respect to
variations in the relevant parameters~\cite{nonfermi}.
Recently, Shankar~\cite{rgshankar} and Polchinski~\cite{polch}
have presented an intuitively appealing idea
of using the Renormalization-Group (RG) approach, so powerful
in particle and statistical physics, to systems of interacting
electrons with a Fermi surface in order to understand, at least
qualitatively, how deviations from Fermi liquid
behaviour can appear {\it naturally} (as opposed to being
fine-tuned).
From this point of view Landau's fermi liquid
is nothing else but a system of free electrons, which has
no relevant perturbations, in the RG sense,
that can drive it away from its trivial infrared
fixed point. In general, however,
as we integrate out certain modes of our original theory, some
interactions may become relevant in the RG sense,
i.e. their effective coupling may grow as one lowers
the momentum scale. Then, two interesting possibilities
arise~\cite{polch}. (i) Fermion bound states are formed, symmetries
are spontaneously broken, and the low-energy spectrum bears little resemblance
to that of the original theory. In such a case one has to re-write
the effective theory in terms of the new degrees of freedom :
for instance, in the superconducting case this is the
Landau-Ginzburg effective action expressed in terms of the fermion
condensate. (ii) Alternatively, the growth of the coupling
is cut off by quantum
effects
at a certain low energy scale, and in this way a
{\it non-trivial} fixed point structure emerges.
The low energy fluctuations still correspond
to fields of the original theory despite their
non-trivial interactions. This case leads
to observable deviations from
the Fermi-liquid behaviour.
\pr
In the case of the high-$T_c$ materials, the physically
interesting question is whether one model theory
can be found with a structure rich enough
to describe {\it both} the non-fermi liquid behaviour
of the normal phase {\it and} the transition to
(and phenomenology of) the superconducting phase. In this
article we shall put forward a candidate
model which, as we shall argue, seems to us to fulfill this r\^ole.
\pr
It is known that possibility (i) above can be caused by relevant interactions
of superconducting (BCS) or charge-density-wave (CDW) type,
both of which are accompanied by the formation of fermion condensates.
Possibility (ii) has only rather recently begun to be seriously
explored~\cite{rgshankar,polch,nayak}. It has been known for a long time
that the electromagnetic interaction of the vector potential
can cause deviation from fermi-liquid behaviour~\cite{vanalphen},
but its effects are suppressed by terms of $O[(v_F/c)^2]$,
with $v_F$ the Fermi velocity and $c$ the light velocity.
Its effects occur only
at much lower energies than those relevant to the high-$T_c$
materials. Nevertheless, the electromagnetic example
is suggestive enough,
perhaps, to motivate a search for other (non-electromagnetic)
gauge interactions in which the effective
signal velocity would be of order $v_F$, and which
might be responsible for a non-trivial fixed point behaviour.
It was precisely this sort of (``statistical'')
gauge-fermion interaction that was studied
(in different forms) in \cite{polch} and \cite{nayak},
and which led to non-trivial fixed point structure
in the infrared.
\pr
Returning
now to possibility (i), we recall that
it has been shown~\cite{doreym}
that a variant of $QED$ in (2 + 1)-dimensions ($QED_3$) leads
to superconductivity, characterized - as appropriate to two space dimensions -
by the absence of a local order parameter (Kosterlitz-Thouless
mode). Thus the exciting possibility arises that a single fermion-gauge
theory could describe both non-fermi-liquid
behaviour in the normal phase and the transition to the
superconducting phase.
\pr
The main purpose of the
talk
is to review an (approximate) renormalization group
analysis~\cite{aitmav}
of a simplified version
of this model, namely $QED_3$ itself, which indicates
that $QED_3$ exhibits two quite
different behaviours depending on the momentum scale.
At very low momenta $QED_3$
enters a
regime of dynamical mas generation (d.m.g.), which in the full theory
leads to superconductivity;
but at ``intermediate'' momenta (see below) d.m.g. does not
occur and the dynamics
is controlled by a non-trivial
fixed point, leading to non-fermi liquid behaviour. Thus we have
the possibility - for the first time, to our knowledge -of one theory
encompassing both the normal and the superconducting phases of
the high-$T_c$ cuprates.
\pr
At this point the reader might worry that applying
renormalization group techniques to a super-renormalizable
theory like $QED_3$ is redundant, since the theory
has no ultraviolet divergencies.
However, this is a mistaken view.
In the modern approach to the RG and effective field theories, one
considers quite generally how a theory evolves
as one integrates out degrees of freedom above
a certain momentum scale, moving progressively down in scale.
From this point of view an effective field
theory description is equally applicable to non-renormalizable,
renormalizable, and super-renormalizable theories.
However, there are some crucial
new features in the case of a super-renormalizable
theory (which, to our knowledge, have not
been identified hitherto). First, the $QED_3$ coupling $e$
introduces an intrinsic {\it intermediate}
scale $e^2$ which has the dimension of mass, this being directly
related to the super-renormalizability of the theory. The physical
effect of this will be the existence of an intrinsic
mass scale and we can expect different physics in different
regimes of momenta relative to this mass scale ($p >> e^2$,
$p \simeq e^2$, $p << e^2$).
\pr
The second distinctive feature of our RG analysis of $QED_3$,
concerns the
way in which we introduce a running coupling. Conventionally,
such running couplings are dimensionless - so, once again the
dimensionfulness of $e^2$ presents a new feature. The way in
which an effective dimensionless running coupling can be introduced into
$QED_3$ has been shown
by Kondo and Nakatani (KN)~\cite{kondo}, building on work by
Higashijima~\cite{higashijima} for $QCD_4$.
The crucial step is to consider the effect of wavefunction
renormalization in the Schwinger-Dyson (SD) equations, as controlled
by a large-$N$ approximation. In this case, one considers
the theory at large $N$ with $\alpha =e^2 N$ held fixed,
and the dimensionless coupling that runs is essentially
$1/N$.
\pr
KN actually considered only the regime in which dynamical mass generation
(chiral symmetry breaking) occurs - and of course here the gauge coupling
is becoming strong and the use of a large-$N$ expansion
is unavoidable. What we did in ref. \cite{aitmav} is to
identify the ``normal'' (no dynamical mass generation)
regime of the theory, and extend the RG-type analysis of KN to this
normal regime.  We argued that there exists a non-trivial
(quasi-)fixed point
of the effective dimensionless coupling, which governs
the dynamics for a range of {\it intermediate}
momenta $p \simeq \alpha $, lying between
the trivial fixed point at $p >> \alpha $, and the region
$p << \alpha $ of dynamical mass generation.
Important to this analysis will be the introduction
(following KN) of an infrared cutoff $\epsilon $, which
serves to delineate the different momentum regimes.
The analysis of ref. \cite{aitmav}
is performed at zero temperature.
Some attempts have also been made
to connect this to finite-temperature
calculations, by interpreting
the temperature as an effective infrared cutoff. We presented
an approximate computation, at finite temperature, of the
electrical resistivity
$\rho $ of the fermionic system.
We argued that it is the existence of the non-trivial RG fixed point
which is responsible for the fact
that the non-fermi liquid behaviour
($\rho $ approximately proportional to the
temperature $T$) is observed
over so large a temperature range. Wavefunction
renormalization effects, important at $O(1/N)$, lead to
calculable logarithmic deviations from the linear in $T$ behaviour.
\pr
At this stage it is useful
to compare and contrast our approach with two other
recent explorations of gauge theories in $(2 + 1)$ dimensions
in a similar context, by Polchinski~\cite{polch} and by
Nayak and Wilczek~\cite{nayak}. Both works deal with
fermions interacting with a statistical gauge field,
the latter representing magnetic spin-spin
interactions.
In both,
the fermions represent {\it spin} quasi-particle
excitations (spinons), and they should therefore
not be identified with the carriers of ordinary
electric charge (holes or electrons). This is to be
sharply contrasted with our own model of refs.
\cite{doreym,aitmav},
in
which the spin-charge separation is done differently,
leading to the fermions in our model carrying both
statistical and ordinary charge.
\pr
The alert reader might worry at our cavalier use of
a relativistic fermion field theory ($QED_3$)
to infer conclusions pertaining
to complicated condensed matter systems
with non-trivial fermi surfaces, like the ones
of relevance to high-temperature superconductivity.
To such objections, we first stress that the results
of ref. \cite{aitmav} should only be viewed
as a {\it qualitative}
attempt at identifying one particular (but important)
source of (cross-over) deviations from fermi liquid behaviour.
In support of this we refer the reader
to an important observation by Polchinski~\cite{polch}
according to which,
in such condensed matter systems,
kinematics implies that the most important
interactions among fermions
are those which pertain to fermionic excitations
whose momentum components tangent to the fermi surface are
parallel. This is the only way that the gauge field momentum
transfer can still be relatively large as compared to the
distance of the fermion momenta from the fermi surface, as required
by special kinematic conditions~\cite{polch}.
There are two cases where such conditions are met
in condensed matter physics. The first pertains to
nested fermi surfaces, at which the points with momenta
$k_0$ and $-k_0$ have parallel tangents. This is
the situation relevant to  BCS or CDW.
The other situation, which is the bulk of Polchinski's
work and will be of interest to us as well, is the case
where the fermions are close to a single point
on the fermi surface.
This means that the most important fermion interactions
are those which are local on the fermi surface, and hence
{\it qualitatively}  this situation can be extended to
relativistic (Dirac) fermions  as well, since the
dispersion relations become effectively linear~\cite{doreym}.
\pr
It should be stressed that the curvature of the fermi surface
plays also a non-trivial r\^ole in deviations
from fermi liquid behaviour, since any shape distortion
appears as a relevant RG deformation of the model.
However, as already pointed  out
in ref. \cite{polch} the remarkable stability
of the observed non-fermi liquid behaviour
in the normal phase of the high-$T_c$ materials,
which persists up to temperatures of $600~K$,
cannot be explained by deformations
of the fermi surface, as this would require an un-natural
fine tuning.
It is our belief that a dominant r\^ole
in the phenomenon is played by the statistical gauge
interaction among charged holes,
which arguably characterizes
magnetic superconductors~\cite{doreym}. Support for
this conjecture, within the context of $QED_3$ prototypes,
was one of the main
results of ref. \cite{aitmav}.
\pr
Finally, in an attempt to convince
the more skeptical formal readers
about the qualitative validity of the relativistic
models as prototypes for the description of such
phenomena in condensed matter,
we draw his/her attention to the fact that
the quasi-fixed point behaviour that seems to
characterize~\cite{aitmav} $QED_3$ at $T=0$,
seems to persist for a wide range of finite tempratures $T >0$,
where Lorentz invariance is definitely lost.
\pr
Another important point, which was recently pointed out
by Shankar~\cite{rgshankar} in connection with the
RG approach to interacting fermions,
is the use of an effective large-$N$ expansion
in cases where the effective momentum cut-off
$\Lambda $ is much
smaller than the size of the fermi surface   $k_F$,
$\Lambda /k_F \rightarrow 0$.
Such a situation is encountered in
a RG study of (deviations from) fermi liquid theories,
the Landau fermi-liquid theory being defined as a trivial
infrared fixed point in a RG sense.
To understand the connection of a large-$N$ expansion with
infrared behaviour of excitations
one should recall the work of ref. \cite{gallavoti}
where the RG approach to the theory of the Fermi surface
has been studied in a mathematically rigorous way.
The basic observation of ref. \cite{gallavoti} is that,
unlike
the case of relativistic field theories,
in systems with an extended fermi surface,
the fermionic excitation
fields exhibiting the correct scaling are not the original
excitations, $\psi _x$ ($x$ a configuration space variable),
but rather {\it quasiparticle} excitations
defined  as follows :
\be
    \psi _x= \int _{|{\bf \Omega} |=1} d{\bf \Omega} e^{ik_F {\bf \Omega}
. {\bf x}}
\psi _{x,\Omega} = \int _{|{\bf \Omega} |=1} d{\bf \Omega}
e^{i(k_F {\bf \Omega} -{\bf K}). {\bf x} } {\tilde \psi}_{{\bf K,\Omega}}
\label{relgal}
\ee
where for the shake of simplicity we assumed that the fermi surface
is spherical with radius $k_F$,
${\bf \Omega}$ is a set of angular variables defining
the orientation of the momentum vector of the excitation
at a point on the fermi surface, and
the
tilde denotes ordinary Fourier transform in a momentum space ${\bf K}$.
These quasiparticle fields
have propagators with the correct scaling~\cite{gallavoti},
which
allows
ordinary RG techniques, familiar from relativistic
field theories, to be applied, such as the appearance of
renormalized coupling constants, scaling fields etc.
Indeed it is not hard to understand why this is so.
For this purpose it is sufficient to observe
that for large $k_F$
the exponent of the exponential
in (\ref{relgal}) is nothing other
than the {\it linearization},
${\bf k} \equiv {\bf K}- k_F {\bf \Omega }$,
about a point on the fermi surface, which makes
these quasiparticle excitations identifiable
with ordinary field variables of the low-energy limit
of these condensed matter systems. The latter is a well-defined
field theory~\cite{doreym}.
The crucial point in this interpretation is that
now the field variables will depend on `internal
degrees of freedom', ${\bf \Omega}$,
which denote angular orientation of the momentum vectors on the
fermi surface. In two spatial dimensions, which is the case of
interest, ${\bf \Omega}$ is just the polar angle $\theta$.
Following ref. \cite{rgshankar}
we discretize this angular space into small cells of extent
$f(\Lambda/k_F) << 1$, e.g. $f=\Lambda/k_F$:
\be
\int \frac{d^2k}{4\pi^2} \equiv
\int _{-\Lambda}^\Lambda \frac{dk}{2\pi}
\int_{f(\Lambda/k_F)}^{f(\Lambda/k_F)} k_F
\frac{d\theta}{2\pi}
\label{integration}
\ee
where $k$ denotes a linearizing momentum about a point
on the fermi surface.
Doing so, we observe~\cite{rgshankar}
that when looking at interaction
terms involving fermionic particle-antiparticle pairs,
${\overline \psi}\psi$,
the leading interactions are among those fermion-antifermion pairs
for which the creation and anihilation
operators lie within the same
angular cell.
This is for purely kinematic reasons in the infrared regime
$\Lambda << k_F$, similar to those
mentioned previously~\cite{polch},
which implied that the most
important fermion interactions on the fermi surface must
be among excitations which have their
tangents to the fermi surface
parallel. It is, then, straightforward to see
that interaction terms involving either gauge excitations or
just fermions resemble those in large-$N$
relativistic field theories, given that
the only $\Lambda$ dependence appears through proportionality
factors $f(\Lambda/k_F) << 1$
in front of the interactions, in the infrared.
One, then, identifies $1/N$ with $f(\Lambda/k_F) << 1$,
and the only
difference from ordinary particle-physics large-$N$ expansions
is the dependence of this effective $N$ on the cut-off $\Lambda$:
that is to say, $1/N$ runs.
\pr
As we showed in ref. \cite{aitmav}, however,
large $N$ expansions in three dimensional $QED$ can exhibit
such scale dependence.
Wave-function renormalization
leads to a renormalized
`running' $1/N$. Furthermore, the running is of a
novel nature.
Instead of finding a non-trivial
infrared fixed point, we shall demonstrate the
existence of an (intermediate) regime of momenta, where
the effective running of the gauge coupling, which is essentially
$1/N$ times a spontaneously appearing scale, is slowed down
considerably, so that one encounters a quasi-fixed-point
situation. This quasi-fixed point
structure is sufficient to cause (marginal) deviations
from the fermi liquid picture.
At finite temperatures, there are indications~\cite{aitmav}
that this behaviour will lead to logarithmic temperature-dependent
corrections to the linear resistivity
of the fermion
system, the latter being the
result of the presence of (statistical)
gauge interactions.
This makes
such theories plausible candidates for a correct
qualitative description of deviations from
Landau fermi liquid theory, which might be related
to the observed anomalous normal phase properties of
high-$T_c$ cuprates.

\section{$QED_3$: Super-renormalizability,
`running' couplings and non-trivial (quasi-)fixed-point structure}

\pr
Three-dimensional quantum electrodynamics ($QED_3$)
has recently received a
great deal of attention (\cite{app}-\cite{ait})
not only as a result of its potential application to the study of
planar high-temperature superconductivity~\cite{doreym}, mentioned
in the introduction, but also because
of its use as a
protoype for studies of chiral symmetry breaking
in higher-dimensional (non-Abelian) gauge theories~\cite{miran}.
\pr
Despite the theory's apparent simplicity  the
situation is not at all clear at present.
A great deal of controversy has arisen
in connection with the r\^ole of the wave-function
renormalization. In the early papers~\cite{app}
the wave-function renormalization $A(p)$ was argued to be
$1$
in Landau gauge
to leading order in $1/N$, where $N$ is the number of fermion
flavours, and thus was ignored.
More detailed studies, however, showed~\cite{pen}
that the precise form, within the resummed $1/N$
graphs, of $A(p)$ is
\be
     A(p) = (\frac{p}{\alpha })^{\frac{8}{3N\pi ^2 }}
\label{one}
\ee
where $\alpha =e^2 N$ is the dimensionful coupling constant
of $QED_3$, which is kept fixed as $N \rightarrow \infty $.
It is clear from (\ref{one}) that, although at energies
$p \simeq \alpha $ the wave-function is of order one,
however at low momenta $p << \alpha $, relevant for
dynamical generation of mass, the wave-function renormalization
yields logarithmic scaling violations which could affect~\cite{pen}
the
existence of a critical number of flavours $N_c$, below which,
as argued in ref. \cite{app}, dynamical mass generation
occurs~\footnote{However, this result was not free of ambiguities
either, given that the inclusion of wave-function
renormalization necessitates the introduction of a non-trivial
vertex function. The exact expression for the latter is not
tractable, even to order $O(1/N)$, and one has to assume various
ansatzes~\cite{pen} that can be questioned.}.
The situation became clearer after the work of ref. \cite{kondo},
who showed that the introduction of an infrared cut-off
affects the results severely, depending on the various
ansatzes used for the vertex function.
In particular,
there are extra
logarithmic scaling violations in the expression for $N_c$,
depending on the form of the vertex function assumed,
which render the limit where the infrared cut-off is removed,
not well-defined.
\pr
For our present purposes, however, we are not
so much interested in whether the inclusion
of wavefunction renormalization leads to a critical
$N_c$ or not, as in the more general point that - as
noted by Kondo and Nakatani (KN)~\cite{kondo}, following
Higashijima~\cite{higashijima} - the vacuum polarization
contribution to $A$ produces effectively
a running coupling, even in the case of the
super-renormalizable theory of $QED_3$. KN's analysis
was restricted to the regime of dynamical
mass generation, and our main purpose in this section
is to extend that to the ``normal'' regime where mass is not
dynamically generated. We emphasize now, however,
that if $A$ is set equal to unity at the outset, the power
of the running coupling concept to unify both regimes
is completely lost.
\pr
We now proceed to a brief review of our analysis in ref.
\cite{aitmav}.
Following ref. \cite{kondo}, we make
the vertex ansatz
\be
\Gamma _\mu (q,p) = \gamma _\mu A(p)^n \equiv \gamma _\mu G(p^2)
\label{three}
\ee
where $p$ denotes the momentum of the photon.
The Pennington and Webb~\cite{pen} ansatz corresponds to
$n=1$, where chiral symmetry breaking occurs for
arbitrarily large $N$~\cite{pis}. It is this case that
was argued to be consistent with the
Ward identities that follow from gauge invariance~\cite{pen}.
In this paper we shall concentrate on the
generalized
ansatz, with $n \ne 1$,
and in particular we shall discuss its finite temperature
behaviour.
We keep the exponent $n$
arbitrary~\cite{kondo} and discuss qualitatively the
implications of the vertex ansatz for various ranges of the
parameter $n$. As we shall argue below this is crucial for
the low-energy renormalization-group structure of the model.
\pr
Using the ansatz (\ref{three}), one
can analyze the Schwinger-Dyson (SD) equations,
in the various regimes of momenta,
in terms of a running
coupling.
For pedagogical purposes, we first concentrate
on the (infrared) regime of dynamical
mass generation, following~\cite{kondo}. The
(approximate) SD
equation for $A(p)$ is (in Landau gauge)
\be
  A(p) = 1 - \frac{g_0}{3} \int _\epsilon ^\alpha dk
\frac{k A (k) G(k^2) }{k^2 A^2 (k) + B(k^2)}
\{(\frac{k}{p})^3 \theta (p - k) + \theta (k - p) \}
\label{sdir}
\ee
where $g_0 =8/\pi ^2 N$, $N$ is the number of fermion
flavours, and $\epsilon$ is an infrared cutoff.
In the low-momentum
region relevant for dynamical mass generation $p << \alpha $
and the first term in the right-hand-side of (\ref{sdir}),
cubic in $(\frac{k}{p})$,
may be ignored. Then, taking into account that $G(k^2)=A(k)^n$,
and using the bifurcation method in which one ignores the
gap function $B(k)$
in the denominators of the SD equations,
one obtains easily
\be
  A(t)=1-\frac{g_0}{3} \int _t^0 ds A^{n-1} (s)
\label{uvsd}
\ee
which has the solution
\be
 A(t) = (1 + \frac{2-n}{3}g_0 t)^{\frac{1}{2-n}}
\qquad ; \qquad
 t \equiv ln (p/\alpha )
\label{uvsol}
\ee
Substituting to the SD equation for the gap, one then obtains
a `running' coupling~\cite{kondo} in the low momentum region
\be
 g^L =\frac{g_0}{1 + \frac{2-n}{3}g_0 t }
\label{two}
\ee
which, we note, is actually independent of $\epsilon$.
The existence of the dimensionless parameter $g^L$
in $QED_3$ may be associated with the ratio
of the gauge coupling $e^2/\alpha  $, given that in the
large $N$ analysis the
natural dimensionful scale  $\alpha$
has been introduced. Thus, a renormalized
running $N^{-1}$ might be thought of as expressing
`charge' scaling in this super-renormalizable theory.
In particular (\ref{two})
implies that the $\beta$ function
corresponding to  $g^L$ is of `marginal' form
\be
    \beta ^L \equiv -\frac{d g^L}{dt} = \frac{2-n}{3}(g^{L})^2
\qquad .
\label{betaf}
\ee
Thus, depending of the sign of $2 -n $ one might have
{\it marginally}
relevant or irrelevant couplings $g^L \propto e^2/\alpha$.
The first derivative of the
$\beta$ function with respect to the coupling $g^L$
is
\be
    \frac{d}{d g^L} (\beta ^L) =2\frac{2-n}{3}g^L
\label{derbet}
\ee
and since $g^L > 0$ by construction, its sign depends
on the sign of $n-2$.
For $n < 2$  (the marginally relevant case)
the gauge interaction decreases rapidly as one
moves away from low momenta,
and the theory is ``asymptotically free''~\cite{kondo}.
If $n > 2$ (marginally irrelevant), on the other hand, then
$g^L (t)$ tends to zero in the low momentum region, whilst for
$n=2$ the coupling is exactly marginal and one recovers the
results of ref. \cite{app,atkinson} about the existence of a critical
flavour number.
Gauge invariance, in the sense of the Ward-Takahashi identity, seems to
imply~\cite{pen,atkinson}
$n \le 2$ and this is the range we shall explore in this article.
\pr
Our task in ref. \cite{aitmav}
was to extend (\ref{two}) beyond
the region $p << \alpha $. Consider first the true ultraviolet
region $p \rightarrow \infty $. Assuming for the moment
that (\ref{two}) were correct for $p >> \alpha $, one finds
a zero of the $\beta$ function
at the point $t \rightarrow  \infty $, the trivial
fixed point $g^* = 0$,
which is
an ultraviolet fixed point.
However, (\ref{two}) or (\ref{betaf})
are not reliable for the range of momenta
$p >> \alpha $. Both formulas
have been derived in the regime of momenta relevant
to the dynamical mass generation,
$p << \alpha $.
\pr
This being so, do we have an alternative argument
for a trivial ultraviolet fixed point?
The answer is affirmative.
To this end we use the results of ref. \cite{kondoquench}
employing a quenched fermion approximation in large $N$
QED. The result of such an investigation is that
once fermion loops  are ignored,
and hence only tree-level graphs (ladders) are
taken into account, the wave-function renormalization
is rigorously proved to be trivial in the Landau gauge~:
\be
      A(p)^{quenched}  = 1
\label{quenched}
\ee
This result is a consequence of special mathematical
relations of resummed ladder graphs in Schwinger-Dyson
equations. Now in our case, one observes that
in the high-energy regime, $p \rightarrow \infty$,
the $\frac{1}{N}$-resummed gauge-boson
polarization tensor vanishes as
$\Pi (p \rightarrow \infty ) \simeq \alpha /8 p \rightarrow  0 $.
Thus, the situation is similar to the quenched approximation,
which implies the absence of any wave-function
renormalization (\ref{quenched}), and therefore
the vanishing (triviality) of the effective (`running')
coupling constant $g$ in the ultra-violet regime of momenta.
This is in qualitative agreement with the
naive estimate made above, based on the
formulas (\ref{two}), (\ref{betaf}).
\pr
The situation is, therefore, as follows.
The coupling grows from the trivial fixed-point
(ultraviolet regime) where there is no mass-generation,
to stronger values as the momenta become lower.
According to the naive formula (\ref{betaf}),
this coupling grows indefinitely for low momenta
and the perturbation expansion breaks down.
But -to repeat - (\ref{two}) was derived
for the regime $p << \alpha $, and the question
now arises whether nothing new happens
from this regime all the way up to $p \rightarrow \infty $, or whether
there is interesting structure at intermediate scales.
In particular, we might envisage a ``quasi-fixed-point'' situation,
in which $g$ remains more or less stationary around the
value $g (0)$ for a wide range of $t$ below $t=0$, before
commencing to grow rapidly at very low momenta.
\pr
The answer~\cite{aitmav} to the above question turns out
to reside, essentially, in the infrared cutoff $\epsilon$
(which, as we noted above, actually disappeared
from (\ref{two})). The coupling of (\ref{two})
is ``asymptotically free'' (i.e. grows rapidly
in the far infrared)  for $n < 2$, {\it provided}
that the ratio $\alpha / \epsilon $ is large enough
- and in this case dynamical mass generation (d.m.g.)
occurs. To get to the region where d.m.g. does not occur,
we must consider smaller values of $\alpha / \epsilon $,
tending ultimately to unity. This is the region
that will yield the effective
non-trivial fixed point structure.
In this case, $p \simeq \alpha $ and hence
the only allowed
region for the momentum
$k$ in (\ref{sdir}) is $ k \le p$, which now eliminates
the {\it second} term in (\ref{sdir}).
Solving then (\ref{sdir}) in this approximation
(and taking $B = 0$ since d.m.g. does not occur), with
the vertex (\ref{three}), one obtains
\bea
   A(p)&=&1 - \frac{g_0}{3} \int _\epsilon ^ p \frac{dk}{k}
(\frac{k}{p})^3 A^{n-1} (k) = \nn \\
1 -& & \frac{g_0}{3}
      \int _{t_0 - t} ^0             ds e^{3s} A^{n-1} (s)
\label{sdifsol}
\eea
which can be easily solved with the result
\be
   A (t) = ( const + \frac{2-n}{9} g_0 e^{3t_0 -3t})^{\frac{1}{2-n}}
\label{solsd}
\ee
where the $const $ is a positive one and can be found from the
value of the wave function renormalization at $t=
ln(\epsilon/\alpha) \equiv t_0$, namely $A(t_0)
=1$. From
(\ref{sdifsol}) this yields
the value $const  =1 - \frac{2-n}{9}g_0$.
Substituting (\ref{solsd}) back to the gap equation
one obtains a `running' coupling constant in this new
intermediate regime
\be
  g^{I}  \equiv \frac{g_0 e^{3t}}{ (1-\frac{2-n}{9}g_0)
       e^{3t} + \frac{2-n}{9}
g_0 e^{3t_0} }  =
\frac{g_0}{ 1-\frac{2-n}{9}g_0
        + \frac{2-n}{9}
g_0 (\frac{\epsilon}{p})^3 }
\qquad .
\label{renir}
\ee
We note that just as the ``lower scale'' $\epsilon$
disappeared from (\ref{two}), so the ``intermediate scale''
$\alpha $ is absent from (\ref{renir}).
\pr
Let us study the fixed-point structure
of this renormalization-group flow .
To this end, consider the $\beta $ function obtained from
(\ref{renir}) :
\be
 \beta ^I = -dg^I / dt = -3 g^I + \frac{ 3     }{g_0}
(1 - \frac{2-n}{9}g_0)(g^I) ^2
\label{betir}
\ee
Taking into account that $g_0 =\frac{8}{\pi^2 N} $
we observe that the vanishing of $\beta ^I $
occurs not only at $g^I = 0$ but also  at the non-trivial
point
\be
    g^I_* =  \frac{8}{\pi ^2 N} (1 - \frac{2 - n}{9}
\frac{8}{\pi ^2 N} )^{-1}
\label{notriv}
\ee
which indicates the existence of a fixed point
lying at a distance of $O(1/N)$, for $N \rightarrow \infty $,
from the trivial one.
\pr
For what momenta is this fixed point
reached ? Accepting (\ref{renir}) at face value,
the answer would be that it is reached for $p \rightarrow \infty $.
But of course (\ref{renir}) is not valid for $p >> \alpha $,
being appropriate for $\epsilon < p < \alpha $ where
the ratio $\epsilon / \alpha $ is smaller than unity, though not
so very small that $p$ can enter
the region of d.m.g. Referring then to the right hand side
of the second equality in (\ref{renir}), we see that when
$p \simeq \alpha $ the quantity $g^I$ will be very close to $g^I_*$,
differing from it by terms of order $(\epsilon/\alpha )^3\frac{1}{N^2}$,
which is negligible. Indeed, as $p$ moves down to $p \simeq \epsilon$,
$g^I$ arrives at $g_0$, which is still within $(1/N^2)$ of $g^I_*$.
Thus the crucial point is that there is - on the basis of this admittedly
approximate analysis - a significant momentum region
over which the coupling $g^I$
varies very slowly, and we are in a ``quasi-fixed-point'' situation.
In a sense, this slow variation of $g^I$ in the range
$\epsilon < p < \alpha $ (for not too small $\epsilon $)
provides a reconciliation between the normalizations adopted
in the two different approximations (\ref{two}) and (\ref{renir}) -
namely
between $g^L (p=\alpha )= g_0$ and $g^I (p=\epsilon )= g_0$.
\pr
The new fixed point occurs at weak coupling for large $N$.
This is consistent with the interpretation that such a fixed point should
characterize a regime of the theory, as determined by the ratio $\alpha /
\epsilon $, where dynamical mass generation does not occur.
\pr
In summary, then, our analysis in ref.\cite{aitmav}
suggests a significant
modification of the picture presented
by Kondo and Nakatani~\cite{kondo}. Whereas those authors
only considered $\epsilon << \alpha $, which is the regime
of ``asymptotic freedom'' and d.m.g., we have explored
also the region of smaller values of
$\alpha / \epsilon $, and have concluded that here quantum corrections
create a quasi-fixed-point with weak coupling.
{\it Both} regions of $\alpha / \epsilon $ are important
in our application of these results to the cuprates,
as we discuss in some detail in ref. \cite{aitmav},
where we tried to relate
the ``$\epsilon $''of this $QED_3$ with the temperature $T$
of $QED_3$ at finite temperature.
\pr
At this stage, it is worth pointing out
the similarity of the above-demonstrated `slow running' of the
effective gauge coupling $g$ at intermediate scales
with (four-dimensional)
particle physics models of `walking technicolour'
type~\cite{techni}.
Such models pertain to gauge theories with
asymptotic freedom and involve regions of momentum scale
at which effective running couplings move very slowly
with the scale, exactly as  happens in our (asymptotically
free) $QED_3$ case~\footnote{A similarity
of $QED_3$ with walking technicolour
had also been pointed
out previously~\cite{dagotto}, but from a different point of view.
In ref. \cite{dagotto}, a formal analogy
of $QED_3$ with walking technicolour models was noted,
based on the r\^ole
of fermion loops in softening the logarithmic confining gauge potential
to a Coulombic $1/r$ type, in the infrared
regime of momenta. This $1/r$ behaviour of the potential,
and its relevance to
dynamical chiral symmetry breaking, is common in both
theories. The formal analogy between $QED_3$ and walking technicolour
theories is achieved~\cite{dagotto}
by replacing the coupling $g^2$ of the four-dimensional
theory
by $1/N$ of $QED_3$.
However $N$ of ref. \cite{dagotto}
does not vary with the energy scale, since wave-function
renormalization effects have not been discussed in their case.
This is the crucial difference in our case, where there is a
more precise analogy
with walking technicolour theories, due to the slowing-down
of the variation of the `effective'
$N$ (\ref{renir}) with the (intermediate) energy scale.}.
This slow running of the coupling
results in such theories in a significant enhancement
of the size of the fermion condensate. In our case,
such condensates are responsible for an opening of a
superconducting gap, and, therefore, one could associate
the slow running of the coupling at intermediate scales
with the suppression of the coherence length of the superconductor
(inverse  of the fermion condensate) in the phase where dynamical
mass generation occurs. Such a suppression, as compared to the
phonon (BCS) type of superconductivity,
which is an experimentally observed and quite
distinctive feature of the
high-$T_c$ cuprates~\cite{kirtley},
appears then, in the context of the above gauge theory
model~\cite{doreym},
as a natural consequence
of the non-trivial quasi-fixed-point
renormalization group structure. Note that
in ref. \cite{doreym}
the enhancement of the superconducting gap-to-critical-temperature
ratio, as compared to  the standard BCS case, had been attributed
to the super-renormalizability of the theory and the
$T$-independence of quantum corrections, features
which are both associated with the above quasi-fixed-point
(slow running) situation as discussed above.
It is understood, of course, that
before we arrive at definite conclusions about the
actual size of the coherence length in the model,
we should be
able to perform exact calculations by resumming the higher
orders in $1/N$ to see whether these features persist.
At present this is impossible  analytically,
but one could hope for (non-perturbative)
lattice simulations of the above
systems~\cite{doreym,kogut}.
\pr
For completeness,
we would like to compare our results~\cite{aitmav}
to other existing results in the literature concerning
the infrared structure of $QED_3$, and in particular
to the results of
refs. \cite{kondo,appnash,nash}.
In ref. \cite{appnash}, it has been argued, on the basis of
a power-counting analysis, which did not make any use of
the Ward-Takahashi identities, that there is no renormalization
of $N$ to any order in $1/N$,
in the infrared regime of the model.
The arguments were based on the softened Coulombic form of the
gauge-boson propagator in the infrared, as a result of
fermion vacuum polarization: $D_{\mu\nu} \propto
(1/q)(g_{\mu\nu} - (1 - \xi) q_\mu q_\nu /q^2)$, in an arbitrary
$\xi $ gauge, for small momentum transfers $q << \alpha $.
It is worth
noticing that such arguments
appear to apply equally well to Abelian as well as non-Abelian
theories, since in the latter case
non-Abelian three or four gluon interactions
could not contribute to the potential scaling-violating
interactions.
This analysis has been performed without
implementing an infrared cutoff,
due to the infrared finiteness of the (zero-temperature) theory.
In the work of ref. \cite{kondo}, which is applied
to the infrared regime, an infrared cut-off is
introduced, which changes the scaling properties of the gauge-boson
propagator. In this case,
the scale-invariant situation seems to occur only for
the value $n=2$ in the vertex ansatz, which notably does not
satisfy the Ward-Takahashi identities~\cite{pen}.
As we have seen, gauge invariance requires $n=1$, and in that case
there exists a running $N$, at infrared momentum scales,
as well as
a finite
critical flavour number, which however is
infrared cut-off dependent, and diverges in the limit
where the cut-off is removed.
\pr
We can also compare this result with that of
ref. \cite{nash}, which
claims to have proven the gauge invariance
of the critical number of flavours in $QED_3$.
There, a non-local gauge fixing
was used; this mixes orders in $1/N$ expansion, in the sense that
the gap function in SD contains now graphs of $O(1/N^2)$,
whilst the wave-function renormalization still remains of $O(1/N)$.
In contrast, the analysis
of ref. \cite{kondo} remains consistently
at leading order in $1/N$, and in
the Landau gauge. The meaning of the non-local gauge fixing
is not clear
if one stays consistently within an order
by order $1/N$ expansion. Nor does gauge invariance
make complete sense in the presence of an infrared cutoff.
\pr
Thus, the key to
a possible explanation of the discrepancy between the works of
ref. \cite{appnash,nash} and ref. \cite{kondo} seems to be hidden
in the higher orders in the large $N$ expansion, as
well the presence of the infrared cut-off. Notice that a naive removal
of the infrared cut-off might lead to ambiguities, as becomes
clear from the work of ref. \cite{zuk} for finite temperature
field theories, provided that one makes~\cite{aklein} the
(physically sensible ) identification/analogy
of the infrared cut-off with the temperature scale, at least
within a condensed matter effective theory framework.
\pr
Now we come to
our case.  As can be seen by the above discussion,
our results can offer a way out
of the above-mentioned discrepancy.
For us,
the momentum regime of interest is not the infrared one,
where dynamical mass generation occurs, but the intermediate
scale. In this regime, the power-counting arguments
of ref. \cite{appnash} do not apply, since the gauge-boson
propagator does not have a simple Coulombic behaviour.
Thus, the wave-function renormalization effects, that appear
to exist in our, admitedly rough, truncation of the SD equations,
might not be incompatible with the results of ref. \cite{appnash},
pertaining to the existence of a critical flavour number.
From our point
of view, this would mean that,
although there is a (slow) running of an effective $N$,
and thus scale invariance is marginally broken, however,
the running of the coupling
is even more suppressed in the infrared, where strong
quantum effects cut off the increase of the
(asymptotically free) coupling. The infrared cut-off then,
appears as the (spontaneous ?) scale, above which a
slow running of the (asymptotically free) coupling
becomes appreciable. In a condensed-matter-inspired
framework, such a spntaneously appearing scale
makes perfect sense, if one associates the infrared
cut-off with the temperature scale~\cite{aklein}.
For momenta
sightly above the infrared cut-off, then, the situation
of KN~\cite{kondo} seems to be valid. This regime may be
viewed as the
boundary regime for which dynamical mass generation still
can happen.
Below the infrared scale,
which
is a regime that makes perfect sense in an infrared-finite
theory such as $QED_3$, dynamical mass generation
certainly occurs, and
the arguments of ref. \cite{appnash}
apply, leading to an effective cut-off of the increase
of the coupling constant. In this regime, the gauge-boson
propagator assumes a softened Coulombic $1/r$ form,
which has been argued to be important for a
(superconducting) pairing attraction among fermions (holes)
in the model of ref. \cite{doreym}.
Such a situation
was envisaged in
ref. \cite{higashijima} for the case of
chiral symmetry breaking
in four-dimensional $QCD$, which in this way was
dissociated from the confining properties of the theory.
\pr
In the work of KN~\cite{kondo} and ours,
all these issues could be confirmed only if
a more complete analysis
of the SD equations, including higher-order $1/N$ corrections,
is performed.
Whether resummation to all orders in $1/N$
washes out completely
the wave-function renormalization effects at intermediate
momenta, leading to an {\it exactly
marginal} (scale invariant) situation,
or keeps this effect at a RG marginal level,
remains an unresolved issue at present.
On the basis of the above discussion,
one would expect that marginal deviations
from scale invariant behaviour at intermediate momenta,
such as the ones
studied in the present work,
survive higher-order analyses,
but they also lead to a critical
number of flavours,
since the latter is an entity pertaining to
the infrared regime of the theory.
Moreover, for us,
who are interested in
performing the analysis in a condensed-matter rather than
particle-theory
framework,
there is
the issue of the ambiguous infrared
limit of the theory at finite temperatures,
which is by no means a trivial matter~\cite{zuk}.
It seems to us that
all these important questions can only be answered
if proper lattice simulations
of the pertinent systems are performed.
At present, the existing computer facilities
might not be sufficient for such an analysis.
\pr
However, as pointed out in \cite{aitmav},
the slow running
of the coupling constant of the model
at intermediate momentum scales,
if true,
is a desirable effect
from a condensed matter point of view, where
both infrared and ultraviolet cut-offs should be kept.
The wave-function renormalization effects, discussed above,
prove sufficient in leading to a (marginal) deviation
of the theory from the fermi-liquid fixed point.
At finite temperatures, this effect can have observable
consequences, and might be responsible
for the experimentally observed
abnormal normal state properties
of the high-$T_c$ cuprates, the physics of
which the above gauge theories
are believed to simulate.
We stress once again that such effects
would be absent in an
exactly marginal situation, like the one suggested
in ref. \cite{appnash}.

\section{Linear behaviour of the (normal phase) Resistivity
in $QED_3$ with the temperature scale}
\pr
We would like to conclude this talk
by making some remarks on the behaviour
resistivity
of $QED_3$, i.e its response to an externally applied electromagnetic
field.
This requires
connecting
the above picture of the behaviour
of $QED_3$ at zero temperature
to that of the same theory at finite
temperature, $T$. In the absence, again,
of anything like an exact solution
in the $T \ne 0$ case,
approximations (quite possibly severe ones)
will have to be made.
However, the
physical aim is clear:
we want to connect the experimental observation
that the electrical resistivity
in the normal phase of the high-$T_c$
superconductors varies linearly with $T$
over a wide range in $T$ from low
temperatures up to a scale of $600$ $K$,
to the existence of the non-trivial
quasi-fixed-point structure
of $QED_3$ found in the previous section.
Qualitatively, the way we shall make the
connection is to interpret the temperature
in finite-$T$ $QED_3$ as (related to) an
effective infrared cutoff. This will follow
from the form of the gauge boson propagator
for $T > 0$, which we analysed in some detail
in ref. \cite{aitmav},
and we shall not repeat here.
\pr
Our aim in this subsection
is to exhibit non-fermi liquid behaviour of the resistivity,
and associate it with the quasi-fixed-point structure at
intermediate scales revealed in the previous section, via the
qualitative connection $\alpha / \epsilon \sim \sqrt{\alpha / T}$.
The resistivity of the model is found
by first coupling the system
to an external electromagnetic field $A$
and then computing the response of the effective
action
of the system,
obtained after integrating out the (statistical) gauge
boson and fermion
fields,
to a change in $A$.
\pr
In the case at hand, in the model of ref. \cite{doreym}
($\tau_3-QED$)
the effective action of the
electromagnetic field, after integrating out hole and
statistical gauge fields~\footnote{Due to the $\tau _3$
structure, as a
result of the bi-partite lattice structure~\cite{doreym},
there are no cross-terms between the statistical
and the electromagnetic gauge fields to lowest non-trivial
order of a derivative expansion
in the effective action. This implies that in this
model the resistivity is determined by the polarization
tensor of the hole (fermion) loop. On the other hand,
in models
where only a single sublattice is used~\cite{ioffe,larkin},
such cross terms arise, which are responsible - after the
statistical gauge field integration - for the appearance
of a conductivity tensor
proportional to   $\frac{\Pi _F \Pi _B}{\Pi _B + \Pi _F}$,
with $\Pi _{B,F}$ denoting (respectively) polarization tensors
for the boson fields of the $CP^1$ model and for the fermions (holes)
in a resummed $1/N$ framework.
In such a case, the conductivity is determined by the
lowest conductivity among the subsystems~\cite{larkin}.
In condensed-matter systems
of this type, relevant for the physics of the normal
state of the high-$T_c$ cuprates,
it is the bosonic contribution that determines
the total electrical resistivity~\cite{ioffe}.},
assumes the form
\be
   S_{eff} = \int A^\mu (p) \Delta _{\mu\nu} A^\nu (-p)
\qquad ; \qquad \Delta _{\mu\nu} = (\delta _{\mu\nu} -
\frac{p_\mu  p_\nu}{p^2})\frac{1}{p^2 + \Pi }
\label{conduct}
\ee
in a resummed $1/N$ framework, with $\Pi $ the one-loop
polarization tensor due to fermions.
The functional variation of the effective action
with respect to $A$ yields the electric current $j$.
{}From (\ref{conduct}) this is
proportional to the electric field $E(\omega)= \omega A$,
in, say,  the $A_0 =0$ gauge,
with $\omega$  the energy.
In the normal phase of the electron system,
the proportionality tensor, evaluated at zero spatial momentum,
is $\sigma _f \times \omega$,
with $\sigma _f$
the conductivity~\cite{larkin}.
From (\ref{conduct}) then,we have
\be
   \sigma _f = \frac{1}{p^2 + \Pi }|_{{\underline P}=0}
\label{polcond}
\ee
where ${\underline P}$
denotes spatial components of the momentum.
\pr
If the effective action were real, then the temperature ($T$)
dependence of the
resistivity
of the model would be given by the $T$-dependence of the
finite-temperature vacuum polarization of the gauge boson.
Thus, following the estimates of ref. \cite{ait} for the
polarization tensor
in the resummed-$1/N$ framework,
we would have immediately obtained a
linear $T$-dependence
for the resistivity. Such a temperature dependence
would actually be valid~\cite{aitmav} for a wide
range of temperatures above the critical temperature
of dynamical mass generation~\cite{doreym},
due to specific features of the ansatzes involved in the
analysis of ref.
\cite{ait}.
\pr
However, things are not so simple.
As first shown by Landau~\cite{landau},
the analytic structure of the
vacuum polarization graphs entering the effective action
(\ref{conduct}) is such that there are imaginary parts
in a real-time formalism~\cite{weldon}.
These imaginary parts are associated with dissipation
caused by physical processes involving
(on-shell) processes
of the type
{\it fermion } $\rightarrow$ {\it fermion} $+$  {\it gauge
boson}. It turns out that
these constitute the major contributions
to the (microscopic) resistivity~\cite{raizer,lee,ioffe}.
In this picture, the latter is
determined by virtue of the Green-Kubo formula~\cite{Kubo}
in the theory of linear response, and it turns out to be
inversely proportional to the imaginary part
of the two-point function of the ``electric'' current
$j_\mu^\psi ={\overline \psi } \gamma _\mu \psi $, evaluated at
zero spatial
momentum. In our case,
in the leading $1/N$-resummed framework, the two-point
function of the electric current is given by the graph
of fig. 1. Adopting the ansatz
(\ref{three}) for the vertex function, the result
for the current-current correlator is
\be
  <J_\mu (p) J_\nu (-p) > \propto (A(p))^n \Delta _{\mu\nu} (p)
 (A(p))^n
\label{five}
\ee
To compute
the imaginary parts of (\ref{five})
would require a real-time formalism,
taking into account the processes of Landau damping~\cite{zuk},
which are not an easy matter to compute
in resummed $1/N$ approximation, especially in the limit of
zero-momentum
trasfer, relevant for the definition of
resistivity. Indeed, as shown in ref. \cite{zuk},
and mentioned briefly above,
there is a non-analytic structure
of the imaginary parts of the one-loop polarization tensors
appearing in the quantum corrections of the gauge boson propagator.
Such
non-analyticities result in a non-local effective action.
This non-locality persists
upon coupling the system to an {\it external} electromagnetic
field $ A$. Since the resistivity of the system is
defined
as the response of the system to a variation of $A$,
then the Landau processes, which
constitute the major contribution
to the (microscopic) resistivity,
complicate the situation enormously.
At present, only numerical treatment of these non -analyticities
is possible~\cite{zuk,ait}.
\pr
We can circumvent this difficulty, and use only the
real parts of the gauge boson polarization tensor
to estimate the
temperature dependence of the resistivity,
by
making use~\cite{aitmav} of the fact that in ``realistic''
many-body systems~\cite{doreym,lee,ioffe}, believed to be
relevant for a
description of the physics of the cuprates,
there is the phenomenon of spin-charge
separation of the relevant excitations.
According to this picture,
the statistical current (responsible for spin
transport) is opposite to the hole current (electric charge
transport)
and this constraint is implemented by the statistical gauge
field, $a_\mu$, that plays the r\^ole of a Lagrange
multiplier~\cite{ioffe}.
The gauge field, on the other hand, is identified~\cite{aitmav},
for physical (on-shell) processes, with the bosonic current
of the spin excitations. The electric charge is, thus,
transported with a velocity which equals
the propagation velocity $v_F$ of the statistical gauge
fields $a_\mu$ .
In non-trivial vacua, such as the the one
pertaining to our system,
the velocity $v_F$ receives quantum  corrections~\cite{pascua}
from vacuum polarization effects.
In a thermal vacuum such corrections are temperature-($T$-) dependent.
\pr
If we represent the (observable) average of the electric current as
$j_\psi = charge \times v_F$, and use Ohm's law to relate it with an
($T$-independent) externally applied electric field
$E$,  $j_\psi = \sigma  . E$, then one observes
that in this picture
the main
$T$-dependence of the resistivity $\sigma ^{-1}$,
comes from
$v_F$, as a result of (thermal) vacuum polarization
effects~\cite{pascua}.
The result is~\cite{aitmav}
\be
   v_F \propto \frac{Q}{T^{\frac{3}{2}}}  \qquad ; \qquad
   Q \rightarrow \epsilon
\label{veloc}
\ee
Using the association of the momentum infrared
cutoff $Q \simeq \epsilon $ with $\sqrt{\alpha/\beta}
\propto \sqrt{T}$, one gets from (\ref{veloc})
a linear $T$-dependence for $v_F^{-1}$, and thus for the
resistivity $\rho$.
Such a linear $T$ dependence is
a characteristic feature of the gauge
interactions, and, as we shall discuss below,
is valid for a wide range of $T$.
\pr
Incorporating
wavefunction renormalization
effects in the above analysis
one can easily~\cite{aitmav}
demonstrate  the existence of
(logarithmic) deviations
from this linear $T$ behaviour.
This part of the analysis does not require
an explicit computation of the imaginary part
of the correlator (\ref{five}).
It only requires
$A$ evaluated at $p=0$.
The resistivity, which formally
is given by the imaginary part
of the inverse of (\ref{five}) as $p \rightarrow 0$,
turns out~\cite{aitmav} to have
the following temperature dependence (resummed up to $O(1/N)$):
\be
   \rho \propto O(T^{1 - \frac{1}{4N\pi}})
\label{seventeen}
\ee
where we have taken $n=1$ as in \cite{pen}.
We cannot, in any case,
take the precise value of the exponent in (\ref{seventeen})
seriously in view of the rough approximations made along
the way.
\pr
 However the region $\beta \alpha >> 1$
is, in fact, that of dynamical mass generation, rather than
the ``intermediate'' region $\beta \alpha \gsim 1$ in which
we expect the quasi-fixed-point structure to play a r\^ole.
A numerical analysis shows~\cite{aitmav}
that for a wide range of temperature
below $\alpha$, but not so low that the
symmetry-breaking phase is entered, the resistivity
should have the form (\ref{seventeen}), where the precise
coefficient of the $1/N$ power is not known accurately from the
above analysis.
The main point, then, is the ``stability'' of this $T$-dependence
which correlates remarkably with the quasi-fixed-point structure
discussed above.

\pr
\section{Conclusions and Outlook}
\pr
In this talk we have reviewed results of some
recent work~\cite{aitmav}, which provide evidence for
certain interesting
effects of the wave-function renormalization in (a
variant of)
$QED_3$ that is believed to be a qualitatively correct
continuum limit of semi-realistic condensed
matter (planar) systems simulating high-temperature
superconducting cuprates.
\pr
Based on an (approximate)
Schwinger-Dyson (SD) improved Renormalization Group (RG)
analysis, we have argued for the existence of an (intermediate)
regime of momenta, where the running of the
renormalized dimensionless
coupling of multiflavour
$QED_3$, which is nothing other than the inverse of the
flavour number, is considerably slowed down, exhibiting
a behaviour similar to that of `walking technicolour' models of
particle physics. This slow running, or (quasi) fixed point
structure, has been argued to be responsible
for an increase of the chiral-symmetry breaking
(superconducting) fermion condensate
of the model, as well as for a (marginal) deviation from the
Landau fermi-liquid fixed point. In connection with the latter
property,
we have argued that the large $N$ expansion is fully justified
from a rather rigorous renormalization group approach to
low-energy interacting
fermionic systems with large fermi surfaces.
Some experimentally
observable consequences of this (marginal) non-fermi liquid
behaviour, including
logarithmic temperature-dependent corrections to the
linear resistivity, have been pointed out, which could be
relevant for an explanation of the abnormal normal-state
properties of the high-$T_c$ cuprates.
\pr
The above RG-SD analysis was, however,
only approximately performed at present.
To fully justify the above considerations, and to make
sure that the above-mentioned
effects are not washed out in an exact treatment, one
has to perform lattice simulations of the above models.
Given that this might not be feasible yet, due to the
restricted capacities of the existing computer devices,
an intermediate step would be to perform
a more complete analytic RG treatment
of the relevant large-$N$ SD equations at finite temperatures.
Such a treatment is not easy,
however, due to the mathematical complexity
of the involved equations. In addition, finite-temperature
field theory is known to exhibit unresolved
ambiguities concerning the low momentum limit,
which complicates the situation.
Some of these issues constitute the object of intensive
research effort of our group at present, and we hope to be able
to reach some useful conclusions soon.
\pr
Our work made use of relativistic fermion systems.
We have provided evidence that this might capture
the correct qualitative features responsible
for the observed deviation from fermi liquid behaviour
in realistic high-temperature superconducting systems,
which are known to be characterized by large fermi surfaces.
Indeed, the remarkable stability in the observed behaviour
up to temperatures of $600~K$ cannot be ascribed to
simple
deformations of the fermi surface, which would require an
unnatural fine tuning.
The presence of gauge interactions, of the type considered
in this work, with subtle wave-function renormalization
properties, provides a natural and simple explanation
of the phenomena
in terms
of a (quasi)-fixed point (i.e. cross over) behaviour, rather
than a new universality class. This should be contrasted to
the works of refs. \cite{nayak,polch}, where the existence of a fixed
point was argued.
This is a non-trivial point to have in
mind for possible experimental searches in the future.
Of course, it is understood that
in order to explain the complete set of the
observed properties
of the normal phase of high-temperature
superconductors
the simple relativistic $QED_3$ picture advocated above
is not sufficient, not even qualitatively. One should
probably take into account all possible sources of deviation,
including the ones
arising from the curvature of the fermi surface and its distortions,
etc, in order to arrive at a quantitatively satisfactory
picture of the situation.
\pr
In this context it might be worth pointing out that our
results are also of value for cases of condensed matter
systems with relativistic spectra around some nodes of their
fermi surfaces.
At present, we do not have a physical intutition on the
microscopic
nature
of the gauge interactions that might be involved
in such situations, neither
are we aware of realistic candidate systems that would
realize such scenaria.
A plausible testing ground for these ideas
would be
the case of $\nu=1/2$ fractional
quantum Hall systems~\footnote{We thank
A. Tsvelik for a discussion on this point.}. Experimentally,
there appear to be
deviations from fermi liquid behaviour in such systems,
and there are
recent theoretical
attempts~\cite{nayak2} to relate this to the existence of new
infrared fixed points. From our point of view, if
the $\nu = 1/2$ Hall case is to be characterized by a new
gauge interaction due to, say, interactions among
the magnetic moments of the (planar) electrons,
then our work shows~\cite{aitmav} that it is more likely to
be characterized by
a cross-over behaviour rather,
than the appearence of a non-trivial infrared fixed point.
We hope to study these fundamental problems in the future.
\pr
Closing, we would like to stress once again the exciting
atmosphere for collaboration between
particle-physics and condensed-matter
communities triggered
by the discovery of
fractional quantum Hall systems
and
high-temperature superconductors.
Indeed, as we have heard at this meeting~\cite{laughlin},
there is a plethora of striking
resemblances between many phenomena that characterize these
solid state
systems with the corresponding phenomena in particle physics.
The very nature of the spin-charge separation, which is
essential for magnetic scenaria of high-temperature
superconductivity, seems to be analogous to the
quark fractional charge phenomenon inside the hadron~\cite{laughlin}.
This `constituent-fermion' picture of the
planar holes  in magnetic superconductor models,
 which was
also extended recently to Hall systems as an
attempt at
a {\it microscopic} understanding of the fractional quantum Hall
effect~\cite{greiter}, is strongly reminiscent
of the quark model of hadrons.
This is not unrelated to the gauge approach to the
high-$T_c$ problem advocated in refs. \cite{doreym} and \cite{aitmav},
and briefly discussed above,
where the asymptotic freedom
of the abelian three dimensional gauge field
plays a crucial
r\^ole in determining
the infrared behaviour of the system in connection with
either dynamical mass generation,
related to the superconducting phase, or with the anomalous
properties of
the normal phase.
The situation is analogous
to chiral symmetry breaking in four dimensional QCD.
In this context, the reader's attention is drawn to
recent
numerical evidence~\cite{laughlin} for a QCD-like-string (hadronic)
Regge-pole strucure
in the physical spectrum
of Hubbard or $t-j$ models, which was associated with the
spin-charge separation property.
Certainly this line of research
appears very interesting and exciting, and should be pursued
further.

\section*{Acknowledgements}
\pr
N.E.M. wishes to thank A. Devoto and the
members of the organizing committee of the
Fourth Chia Meeting on {\it Common Trends in
Condensed Matter and High Energy
Physics}, Chia Laguna, Sardegna (Italy),
3-10 September 1995, for the opportunity they gave him to
present results of this work, and for creating
a very stimulating atmosphere during the meeting.
N.E.M also acknowledges informative discussions with A. Barone,
M.C. Diamantini, G. Semenoff, P. Sodano and C. Trugenberger.
N.E.M. thanks P.P.A.R.C. (UK) for an Advanced Fellowship.

 \end{document}